# Synthesis, crystal structure, site occupancy and magnetic properties of aluminum substituted M-type Sr hexaferrite SrFe$_{12-x}$Al$_x$O$_{19}$ nanoparticles.

Matilde Saura-Múzquiz,[a,†,*] Anna Zink Eikeland,[b,†] Marian Stingaciu,[b] Henrik Lyder Andersen,[c] Maxim Avdeev,[d] and Mogens Christensen[b]

[a] Departamento de Física de Materiales, Faculty of Physics, Universidad Complutense de Madrid, 28040 Madrid, Spain
[b] Center for Materials Crystallography, Department of Chemistry and iNANO, Aarhus University, Langelandsgade 140, DK-8000 Aarhus C, Denmark
[c] Instituto de Ciencia de Materiales de Madrid (ICMM), CSIC, 28049 Madrid, Spain
[d] Australian Nuclear Science and Technology Organisation (ANSTO), New Illawarra Road, Lucas Heights NSW 2234 Australia.

[†] Equally contributing authors

*Corresponding author: M. S.-M.: matsaura@ucm.es*

## Abstract

The synthesis of aluminum substituted strontium hexaferrite nanoparticles (SrFe$_{12-x}$Al$_x$O$_{19}$ with $x$ = 0-3), *via* three different preparation methods, is investigated. The synthesis methods are hydrothermal autoclave (AC) synthesis, a citrate sol-gel (SG) synthesis and a solid-salt matrix (SSM) sol-gel synthesis. Evaluation of macroscopic magnetic properties and of lattice parameters obtained by Rietveld analysis of powder X-ray diffraction (PXRD) data indicate that successful substitution of Al into the crystal structure is only achieved for the SG method. For the SG sample with $x$ = 3, the coercivity was found to increase by 73% to 830 kA/m, while the saturation magnetization was reduced by 68% to 22.6 Am$^2$/kg compared to the non-substituted $x$ = 0 SG sample. The SSM and AC samples did not show any significant changes in their magnetic properties. To examine the nature of the Al insertion in detail, neutron powder diffraction (NPD) data were collected on the SSM and SG samples. Combined Rietveld refinements of the PXRD and NPD data confirm that effective substitution of the Al ions is only achieved for the SG sample and reveal that Al occupies mainly the $(2a)_{Oh}$ and $(12k)_{Oh}$ sites and to a lesser extent the $(4e)_{BP}$, $(4f)_{Oh}$ and $(4f)_{Td}$ sites. Moreover, the relative degree of site occupation varies with increasing Al substitution. The intrinsic magnetization according to the refined atomic magnetic moments and Al site occupation fractions was extracted from the NPD data and compared with the measured macroscopic magnetization. A remarkable agreement exists between the two, confirming the robustness and accuracy of the Rietveld analysis.

## Keywords

## Introduction

Rare-earth free permanent magnetic materials play a key role in many technological fields. M-type hexaferrite $SrFe_{12}O_{19}$ is a highly used rare-earth free permanent magnetic material due to its relatively low cost, high chemical stability, good magnetic properties and high Curie temperature ($T_C$ = 737 K).[1] Traditionally, $SrFe_{12}O_{19}$ is synthesized by a standard ceramic method.[2] However, new synthesis methods are currently being investigated in order to exploit the potential of reducing the particle size to the nanoscale regime as a means to enhancing the magnetic properties of the material. Some of these synthesis methods include co-precipitation methods,[3] sol-gel methods,[4,5] hydrothermal methods,[6-8] microwave induced combustion methods,[9] and the reverse micelle technique.[10] The produced $SrFe_{12}O_{19}$ samples show remarkably different particle shapes and sizes depending on the synthesis method used, spanning from agglomerated isotropic particles to plates, needles, or tiny super-paramagnetic crystallites. The resulting magnetic properties are highly dependent on the crystallite size,[11] size distribution, morphology[6,12] and texture.[13] To control these characteristics, a carefully designed synthesis route is essential.

A different approach to tailoring the magnetic properties of a material is by modifying its crystal structure, *i.e.*, modifying it at the atomic level. Several studies on the partial cation substitution of M-type ferrites have previously been published, *e.g.*, the substitution of Fe by $La^{3+}$, $Pr^{3+}$ or $Nd^{3+}$ by J. F. Wang *et al.*,[14,15,16] which gave rise to a moderate enhancement of the coercivity, $H_c$. The most significant enhancement of coercivity to date by substitution of a single cation has been achieved by $Al^{3+}$ substitution.[17-19] Using an auto-combustion synthesis method for preparation of $SrFe_{12-x}Al_xO_{19}$ followed by heat treatment, an $H_c$ of 18.1 kOe (1440.3 kA/m) was achieved for $x = 4$ by Luo *et al*.[17] A similar value of 17.75 kOe (1398.2 kA/m) was obtained by H. Z. Wang *et al.*, also for a substitution of $x = 4$, by a glycin-nitrate method and subsequent annealing.[18] However, the highest reported coercivity value for a ferrite material reported to date was recently published by Gorbachev *et al*. for Ca and Al substituted $SrFe_{12}O_{19}$, ($Sr_{1-x/12}Ca_{x/12}Fe_{12-x}Al_xO_{19}$ with x = 5.5) reaching the impressive value of 36 kOe (2871 kA/m) at room temperature, and 42 kOe (3342 kA/m) at 180 K.[20,21,22] These values are extraordinarily high, largely exceeding even the intrinsic coercivity values of the $Nd_2Fe_{17}B$ magnets (~10 kOe; 796 kA/m).[23] However, the enhancement of coercivity diminishes the saturation magnetization, which is often reduced by 50% or more compared to non-substituted $SrFe_{12}O_{19}$. This can be partly explained by the inverse relation between coercivity and saturation magnetisation ($H_c < 2K_1/\mu_0 M_s$), where $K_1$ is the magnetocrystalline anisotropy constant. Given that $SrFe_{12}O_{19}$ adopts a ferrimagnetic structure, with most of the $Fe^{3+}$ sites having spin up and two sites having spin down (illustrated in blue polyhedral in Figure 1), it is assumed that the substituted aluminum occupies the up-spin (predominant spin direction) sites in the hexaferrite structure thereby causing the reduced magnetization. First-principle total-energy calculations based on density functional theory were carried out by V. Dixit *et al.* to study the site occupancy and magnetic properties of Al substituted $SrFe_{12}O_{19}$, with $x = 0.5$ and $x = 1$.[24] The calculations indicate that the Al atoms preferentially occupy the $(2a)_{Oh}$ and $(12k)_{Oh}$ sites, with increasing probability of the $(12k)_{Oh}$ site being occupied by Al as the synthesis temperature increases. However, until now no experimental evidence had been reported to confirm this.

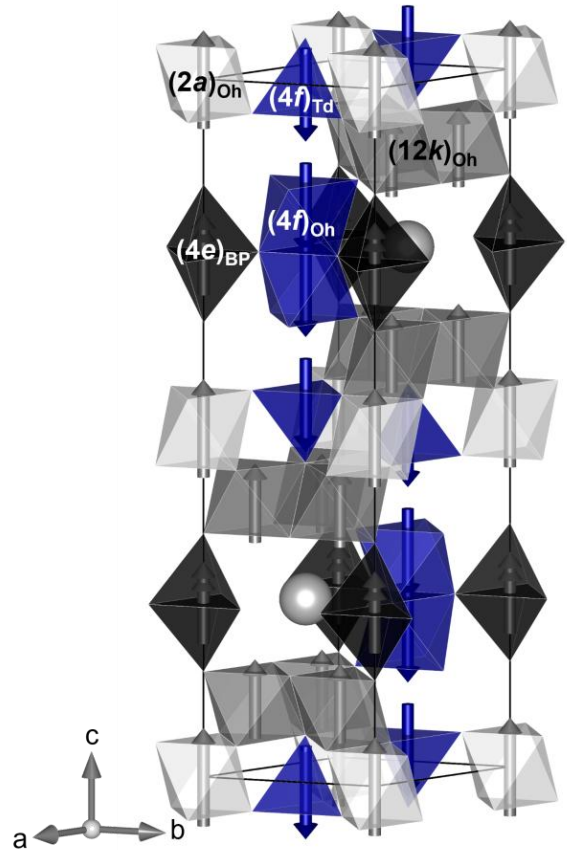

Figure 1: Schematic representation of one unit cell of $SrFe_{12}O_{19}$. The spin-up Fe sites are represented in gray-scale polyhedra, and the spin down sites in blue polyhedral. The magnetic spins are represented by the arrows. The Sr atoms are represented by the grey spheres. The oxygen atoms (not shown) are positioned at the corners of the polyhedra. Illustration made using the software VESTA.[25]

In this study, the syntheses of Al-substituted Sr-ferrite (SrFe$_{12-x}$Al$_x$O$_{19}$, $x = 0–3$) using three different synthesis methods, namely sol-gel (SG), solid-salt-matrix (SSM) and autoclave (AC), have been investigated. The composition, crystal structure and microstructure of the obtained samples are investigated by Rietveld analysis of powder X-ray diffraction (PXRD) data. The magnetic properties of the samples are analyzed by vibrating sample magnetometry (VSM). The data indicates that successful substitution is only achieved using the SG synthesis method. To further investigate the nature and effect of the Al insertion into the structure, joint Rietveld refinements were performed on PXRD and neutron powder diffraction (NPD) data collected on both the SSM and SG samples. This data allowed us to determine the preferred substitution sites as function of aluminum content and to understand its influence on the magnetic properties of the compound.

## Experimental methods

*Sample preparation*

Aluminum-substituted SrFe$_{12}$O$_{19}$ (SrM) nanocrystalline samples were synthesized by three different methods: citrate sol-gel (SG), solid-salt matrix (SSM) and hydrothermal autoclave (AC) synthesis, following similar synthesis procedures to those previously reported by Eikeland *et al.*[4, 7], using starting chemical reagents with purity ≥98% (Sigma-Aldrich or Chem-Solution GmbH, technical grade).

*Sol-gel synthesis (SG)*

Sol-gel synthesized SrM was prepared using Fe(NO$_3$)$_3$·9H$_2$O, Sr(NO$_3$)$_2$, and Al(NO$_3$)$_3$·9H$_2$O powders. The powders were dissolved in a small amount of distilled water under stirring. Dissolved citric acid was added to the solution after which the solution was neutralized using concentrated NH$_4$OH. The molar ratio of [Fe$^{3+}$]:[Sr$^{2+}$] was 11.5, and the nitrate to citric acid ratio was 1:1.

The solution was dried on a hot plate at 100 °C and subsequently heated to 250 °C to let the gel undergo an auto-combustion reaction forming a low-density grey powder. Finally, the precursor was calcined at 925 °C for 30 minutes. The [Fe$^{3+}$]+[Al$^{3+}$]:[Sr$^{2+}$] molar ratio was fixed at 11.5.

*Solid-salt-matrix (SSM)*

The solid-salt-matrix synthesis utilizes a solid matrix-based synthesis, where free standing nanocrystallites grow in a NaCl matrix. SrM nanocrystallites were prepared by dissolving SrCl$_2$·6H$_2$O, FeCl$_3$·6H$_2$O, and Al(NO$_3$)$_3$ in distilled water to obtain a [Fe$^{3+}$]+[Al$^{3+}$]:[Sr$^{2+}$] molar ratio of 11. The metal ion solution was then added to a 1 M solution of Na$_2$CO$_3$ under constant stirring. The synthesis was carried out with 0.5 moles of Na$_2$CO$_3$ per mole of Cl$^-$, resulting in a 1:1 molar ratio between Na$^+$ and Cl$^-$. A 5.5 M solution of citric acid was finally added to the mixture.

The molar ratio between citric acid and $Na_2CO_3$ was 1.5:1. The mixture was dried to a gel in a convection oven at 120 °C overnight. The dried gel was crushed in a mortar and placed in a thin layer of approx. 2 mm in a convection oven at 450 °C for 1 hour to burn off the organic residues. Afterwards, the precursor was calcined at 790 °C for 1 hour, followed by cooling to room temperature. The product was washed in water and 4 M $HNO_3$ to remove the NaCl matrix and finally dried at 80 °C.

*Autoclave synthesis (AC)*

Aqueous solutions of $Sr(NO_3)_2$, $Fe(NO_3)_3 \cdot 9H_2O$, and $Al(NO_3)_3 \cdot 9H_2O$ (1 M each) were mixed in the Teflon insert of a steel autoclave. 16 M NaOH was added slowly and dropwise to the mixture under intense stirring. A 50% excess of NaOH compared to the molar amount of $NO_3^-$ was added. The $[Fe^{3+}]:[Sr^{2+}]$ molar ratio was 4 and the $Sr^{2+}$ concentration of the final precursor was 0.1 M. The molar ratio of $[Fe^{3+}]+[Al^{3+}]:[Sr^{2+}]$ was fixed to 4. The 175 mL Teflon lined autoclaves used as reaction chambers were inserted in a preheated convection oven at 240 °C for four hours. The products were washed with water and 4 M $HNO_3$ to wash away $SrCO_3$.

*Characterization*

Powder X-ray diffraction data were collected on the as-synthesized powder samples using a Rigaku SmartLab diffractometer (Rigaku, Japan) equipped with a D/teX Ultra Si-strip detector, Cross Beam Optics (CBO) and Bragg Brentano geometry. A Co $K_\alpha$ rotating anode source was used for the SG and SSM series, whereas the PXRD data of the AC samples were collected using a Cu $K_\alpha$ rotating anode source.

Neutron powder diffraction data were collected on the SG and SSM samples using a wavelength of 2.44 Å at ECHIDNA, the high-resolution powder diffractometer at the OPAL reactor (Australian Nuclear Science and Technology Organisation, ANSTO, Lucas Heights, New South Wales, Australia).[26]

Rietveld analysis of the diffraction data was carried out using the FullProf Suite software package.[27] The Thompson-Cox-Hastings formulation of the pseudo-Voigt function was used to model the peak profile,[28] and Finger's model was used to describe the peak asymmetry due to axial divergence in terms of finite sample and detector sizes.[29] The instrumental contribution to the peak broadening was accounted for by performing LeBail fits of a NIST $LaB_6$ 660b standard[30] measured under the same conditions as the studied samples. From these $LaB_6$ fits, an instrumental resolution file was created and implemented in the refinements of the studied materials to de-convolute the instrumental and sample contributions to the peak width. For the Rietveld analysis, the zero shift, unit cell parameters, atomic positions and isotropic atomic displacement parameters were refined, as well as the Lorentzian refinable size (Y) and strain (X) parameters. The $S_z$ size parameter was also refined to model the anisotropic shape of the crystallites employing the $F(S_z)$ function according to the Platelet Vector Size model as implemented in FullProf.[31] The platelet vector, **k**, was defined as (001). For Al-substituted samples, the atomic site occupation fractions were also refined. Details on the refinement of site occupation fractions of the Al-substituted samples are given in the results and discussion.

The macroscopic magnetic properties of the samples were measured using a Quantum Design Physical Property Measurement System (PPMS) equipped with a Vibrating Sample Magnetometer (VSM). For the magnetization measurements of the powder samples, approximately 10-15 mg of sample were pressed into pellets of 3 mm in diameter using a hand-held press. The specimen was then placed in a tubular brass sample holder, tightly fit between two quartz rods, and fixed in place with Kapton tape, with the surface of the pellet perpendicular to the applied magnetic field.

## Results and Discussion

*Crystal structure and crystallite size/morphology*

Figure 2 shows representative PXRD patterns and Rietveld refinements for the AC, SSM and SG samples with nominal Al-contents of $x=2$. PXRD data and Rietveld refinements of the remaining samples can be found in Supporting Information. The Rietveld analyses confirm the formation of the hexagonal M-type hexaferrite structure (space group $P6_3/mmc$, #194) in all three series (SG, SSM and AC) and across the entire substitution range. Notably, the AC samples with $x = 0$ and $x = 0.5$ nominal Al-contents were found to contain minor quantities of hematite impurity ($\alpha$-$Fe_2O_3$), amounting to 1.9(1) wt.% and 2.2(1) wt.%, respectively. This impurity was also present in the SG samples in very small amounts (1.2(2) wt.% in $x = 1$, 1.9(2) wt.% in $x = 2$ and 1.9(2) wt.% in $x = 3$).

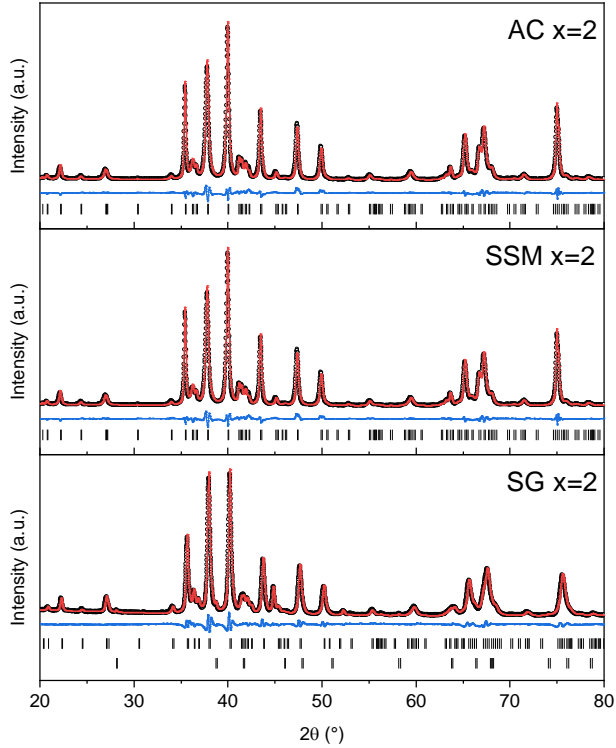

Figure 2: PXRD data (black dots) and refinement model (red line) of $SrFe_{12-x}Al_xO_{19}$ samples with $x = 2$ synthesized by AC, SSM and SG. The difference between the observed and calculated intensities is given in blue, and the black bars mark the position of the Bragg reflections. Note that when NPD data was available (SSM and SG samples), a combined refinement of the structural model to the PXRD and NPD was carried out, although only the PXRD data is shown here.

As demonstrated in previously reported studies,[4, 7, 13] the un-doped $SrFe_{12}O_{19}$ samples exhibit differences in crystallite size and platelet aspect ratio depending on the synthesis method and reaction conditions employed. In the present study, the crystallite size and morphology have been extracted from Rietveld refinement of powder diffraction data using the platelet-vector-size model, as described in the experimental section. The hydrothermal AC method is found to yield relatively thin platelet-shaped crystallites with a thickness, $D_c$, of 41.0(2) nm and diameter, $D_a$, of 154(2) nm. In contrast, the $SrFe_{12}O_{19}$ crystallites prepared by the SG and SSM methods exhibit a more isotropic morphology with diameters of 73.5(5) and 59.8(2) nm and thicknesses of 63(1) and 23.9(1) nm, respectively.

Figure 3 shows the refined crystallite sizes along the *a*- and *c*-axes respectively as function of nominal Al-content for the three series (see Table 1 for numeric values). In the AC series, no clear trend in crystallite size is observed with increasing Al content, although an overall decrease in size along the *a*-axis ($D_a$) can be seen from 154(2) nm at x=0 to 106.3(9) nm at $x = 3$. However, these sizes are very close to the resolution limit of the diffractometer, so the obtained sizes and trends in the $D_a$ of AC samples should not be over-interpreted. In the SSM series, no significant change in crystallite size is observed with increasing Al-content along any of the axes of the platelets ($D_a$ ~50-60 nm, $D_c$ ~22-24 nm). In the case of the SG samples, an initial decrease in crystallite size is observed for the $x = 1$ sample ($D_a$ =61(1) nm, $D_c$ =51(1) nm). However, higher levels of Al-content lead to an increase in crystallite size along both axes that is particularly pronounced along the *c*-axis (for $x = 3$, $D_a$ =86(6) nm, $D_c$ =85(7) nm), as illustrated in Figure 3.

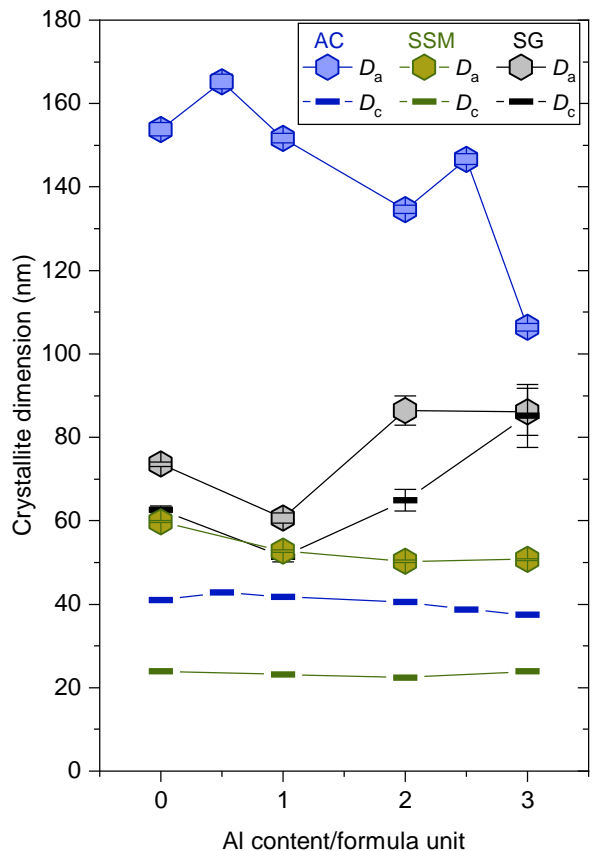

Figure 3: Variation of crystallite sizes along the crystallographic *a*-axis ($D_a$) and the crystallographic *c*-axis ($D_c$) with increasing aluminum content, for samples synthesized by the three different methods, AC (blue), SSM (green) and SG (grey). Where not shown, the errors are smaller than the size of the symbols.

*Evidence for Al-substitution – Lattice parameters and magnetic properties*

Successful isostructural substitution of Al for Fe in the hexaferrite structure would be expected to cause a Vegard's law-like behavior with a gradual reduction in the lattice parameters as function of increasing Al content, due to the smaller effective ionic radius of $Al^{3+}$ compared to $Fe^{3+}$ in tetrahedral (0.39 *vs* 0.49 Å, ~20% smaller), octahedral (0.535 *vs* 0.645 Å, ~18% smaller) and bipyramidal (0.48 *vs* 0.58 Å, ~20% smaller ) coordination.[32] The unit cell axes of the Al-doped $SrFe_{12}O_{19}$ samples ($SrFe_{12-x}Al_xO_{19}$) of the three series (AC, SSM and SG) extracted from Rietveld refinements of PXRD and NPD data, are shown in Figure 4, and the numerical values are given in Table 1. In the case of the AC series, no notable changes in the unit cell axes are observed with increasing Al content. This lack of change suggests that no (or only negligible) effective Al doping is taking place in those samples. In the case of the SSM series, a slight decrease in unit cell axes is observed with increasing Al doping, with the smallest unit cell axes found at compositions of $x = 2$. This small change in unit cell parameters could indicate that some Al insertion is taking place in these samples, although if so, it is very minor. The SG samples, however, exhibit a clear linear decrease of the unit cell axes with increasing Al content, following Vegard's law both along the *a*- and *c*-axes.[33] These results suggest that, unlike the AC and SSM samples, Al is effectively subsituted into the $SrFe_{12}O_{19}$ structure of the SG-synthesized samples.

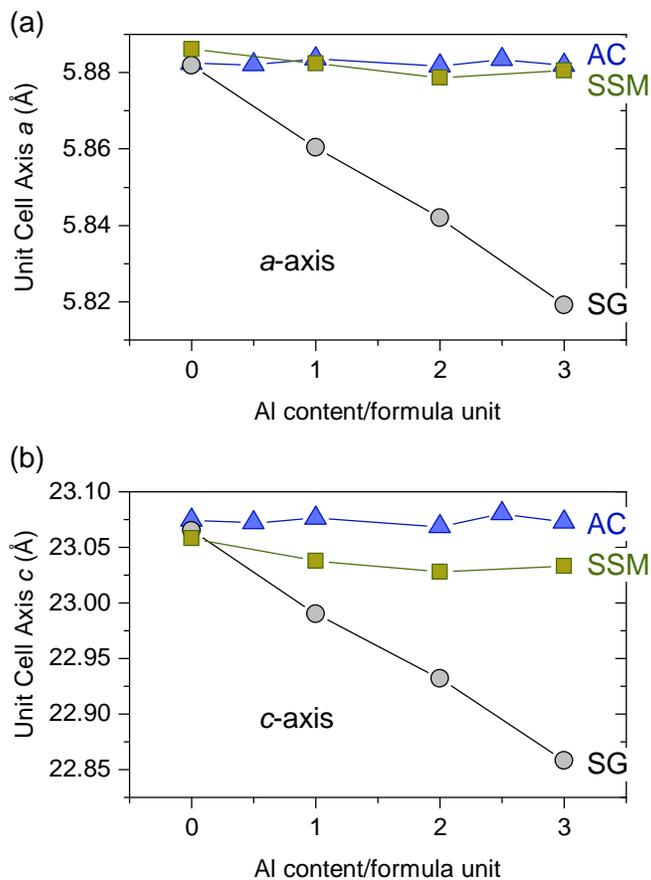

Figure 4: Unit cell (a) *a*-axis and (b) *c*-axis as function of nominal Al-content, extracted from Rietveld analysis of NPD and PXRD data, for samples synthesized by the three different methods; AC (blue), SSM (green) and SG (grey). When not shown, the errors are smaller than the size of the symbols.

Table 1: Refined lattice parameters *a -axis* (Å) and *c-axis* (Å), and crystallite sizes along the *a*-axis, $D_a$ (nm), and *c*-axis, $D_c$ (nm). Values for saturation magnetization, $M_s$ (Am$^2$/kg), remanent magnetization, $M_r$ (Am$^2$/kg), squareness ratio ($M_r/M_s$) and coercive field $H_c$ (kA/m) obtained from VSM measurements.

| $x$ | *a-axis* (Å) | *c-axis* (Å) | $D_a$ (nm) | $D_c$ (nm) | $D_a/D_c$ | $M_s$ (Am$^2$/kg) | $M_r$ (Am$^2$/kg) | $M_r/M_s$ | $H_c$ (kA/m) |
|---|---|---|---|---|---|---|---|---|---|
| AC | | | | | | | | | |
| 0 | 5.88245(2) | 23.0743(1) | 154(2) | 41.0(2) | 3.75(5) | 64.4(2) | 41.5(1) | 0.644(2) | 164(3) |
| 0.5 | 5.88212(2) | 23.0723(1) | 165(2) | 42.8(2) | 3.86(5) | 64.5(1) | 43.43(9) | 0.674(2) | 173(2) |
| 1 | 5.88349(2) | 23.0762(1) | 152(1) | 41.7(2) | 3.64(3) | 67.0(1) | 50.23(9) | 0.750(2) | 143(2) |
| 2 | 5.88166(2) | 23.0684(1) | 134(1) | 40.5(2) | 3.32(3) | 63.1(1) | 44.07(8) | 0.698(2) | 162(2) |
| 2.5 | 5.88338(2) | 23.080(1) | 147(1) | 38.7(2) | 3.88(4) | 63.4(4) | 47.9(1) | 0.754(5) | 125(1) |
| 3 | 5.88186(2) | 23.0728(1) | 106.4(9) | 37.4(2) | 2.84(3) | 63.3(1) | 44.13(9) | 0.697(2) | 144(2) |
| SSM | | | | | | | | | |
| 0 | 5.88617(2) | 23.0579(1) | 59.8(2) | 23.93(9) | 2.50(1) | 70.4(2) | 38.83(7) | 0.551(2) | 482(3) |
| 1 | 5.88241(8) | 23.0377(3) | 52.8(3) | 23.2(2) | 2.27(2) | 61.9(2) | 34.51(6) | 0.557(2) | 549(13) |
| 2 | 5.87875(9) | 23.0278(3) | 50.3(3) | 22.4(2) | 2.25(2) | 62.1(2) | 34.48(5) | 0.555(2) | 568(5) |
| 3 | 5.88050(8) | 23.0334(3) | 50.7(3) | 23.9(2) | 2.12(2) | 66.0(2) | 36.54(6) | 0.554(2) | 519(3) |
| SG | | | | | | | | | |
| 0 | 5.88187(3) | 23.0654(2) | 73.5(5) | 63(1) | 1.17(2) | 71.6(2) | 36.56(4) | 0.511(2) | 479(6) |
| 1 | 5.86033(5) | 22.9904(3) | 61(1) | 51(1) | 1.18(4) | 57(1) | 29.13(6) | 0.51(1) | 492(13) |
| 2 | 5.84190(8) | 22.9321(4) | 86(3) | 65(3) | 1.33(8) | 36.6(3) | 18.37(2) | 0.502(5) | 765(18) |
| 3 | 5.8192(1) | 22.8582(6) | 86(6) | 85(8) | 1.0(1) | 22.6(3) | 11.442(8) | 0.506(6) | 830(21) |

These observations from the refined unit cell parameters are corroborated by the macroscopic magnetic behavior of the samples as illustrated by the field-dependent magnetization curves (mass magnetization, *M*, vs. effective field, $H_{eff}$) shown in Figure 5. Successful substitution of non-magnetic Al$^{3+}$ ions into the structure would be expected to cause a considerable change in the degree of magnetization that can be induced in the samples. The magnetization would be expected to decrease or increase depending on whether Al occupies crystallographic up-spin sites (predominant spin direction) or down-spin sites, respectively (see Figure 1). Here, little-to-no change in the magnetic behavior is observed for both the AC and SSM samples with increased nominal Al content, confirming the lack of (or negligible degree of) Al$^{3+}$ insertion into the structure. In contrast, the SG samples exhibit a clear and gradual change in magnetic behavior with lower magnetization and higher coercivity as the nominal content of Al increases.

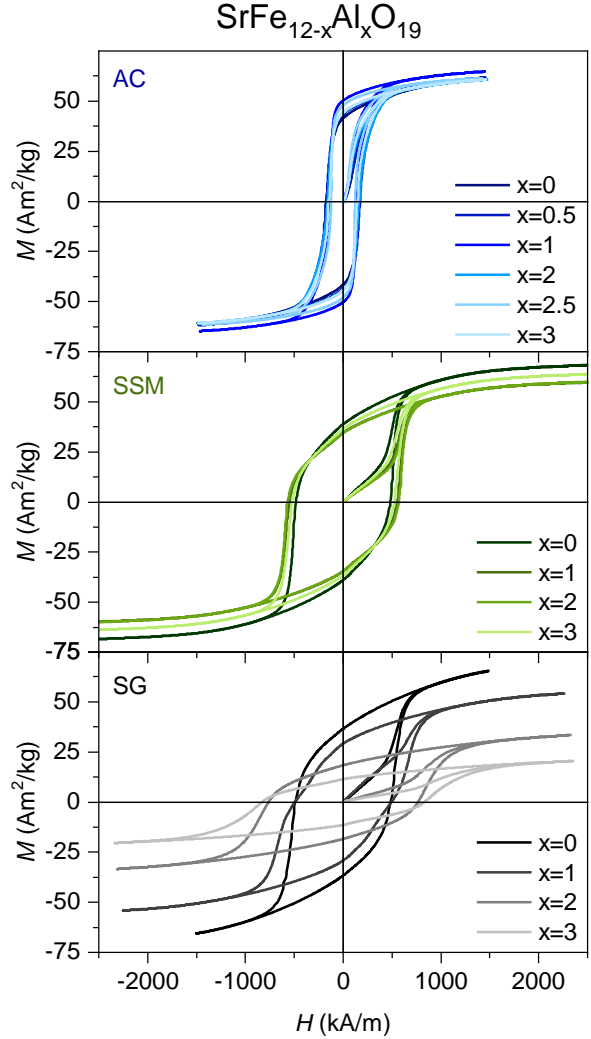

Figure 5: Hysteresis curves (Mass magnetization, $M$, vs. effective field, $H_{eff}$) of the samples from the three different synthesis series (SG, SSM and AC) with different degrees of nominal aluminum substitution.

The extracted values for coercivity ($H_c$), saturation magnetization ($M_s$), and remanence ($M_r$) are shown in Figure 6 and given in Table 1. It should be noted that the saturation magnetization value reported here is not the measured magnetization value at the maximum applied field. Rather, $M_s$ is extracted using the law of approach to saturation,[34] where the measured magnetization of the sample is extrapolated to calculate the true saturation value of the sample. This obtained $M_s$ is sometimes referred to here as "*measured $M_s$*" to distinguish it from the "*calculated $M_s$*" using the refined magnetic moments from NPD data (discussed in the next section).

In the SG series, where effective insertion of the $Al^{3+}$ cation takes place, an enhancement of the coercivity is observed as the Al content increases, from 479(5) kA/m at $x = 0$ to 830(21) kA/m at $x = 3$. The coercivity enhancement is accompanied by a decrease of the saturation magnetization, in accordance with previously reported studies and theoretical calculations,[24] from 71.6(2) Am²/kg at $x = 0$ to 22.6(3) Am²/kg at $x = 3$.

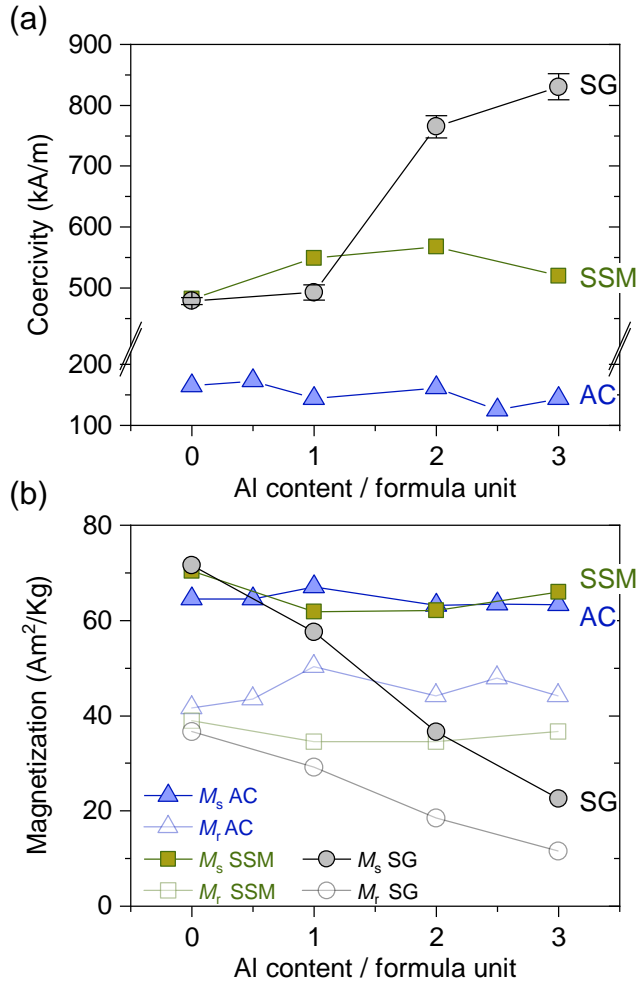

Figure 6: (a) Coercivity and (b) saturation and remanent magnetization of all samples extracted from the measured hysteresis curves, plotted as function of aluminum content per formula unit. When not shown, the errors are smaller than the size of the symbols.

A somewhat similar trend in coercivity and saturation magnetization is observed for the SSM series from $x = 0$ to $x = 2$. The changes in magnetic properties with increasing Al content are, however, much less pronounced than that of the SG samples. Nevertheless, these results could indicate that a very small amount of Al is present in the hexaferrite structure of the SSM samples, although it seems to reach a maximum degree of substitution at $x = 2$, after which the coercivity decreases and the saturation increases, reaching values closer to that of the undoped sample. These results are consistent with the refined unit cell parameters of the SSM samples, where a small decrease in both the $a$- and $c$-axes were observed, attaining the smallest unit cell at $x = 2$. For the AC samples, no considerable change in the magnetic properties is observed with increasing Al-content, confirming once again the lack of successful substitution of Al into the structure.

In addition to the trends in magnetic properties within each series, the data in Table 1 show a clear correlation between the aspect ratio of the nanoparticles ($D_a/D_c$) and the attained $M_r/M_s$. The highest $M_r/M_s$ ratio (0.70) is reached for the AC series, where the more anisotropic particles are obtained, with an average $D_a/D_c$ of 3.53. This is followed by a $M_r/M_s$ of 0.55 for the SSM series, of average $D_a/D_c$ of 2.29, and the smallest $M_r/M_s$ of 0.51 is observed in the SG series, of $D_a/D_c$ equal to 1.17. This result is consistent with our previous studies of hexaferrite nanoparticles, and the influence of morphology in in magnetic texture.[6, 13]

*Aluminum site occupancy and magnetic structure– Combined NPD and PXRD Rietveld refinements*

In order to examine the nature of the Al insertion and its influence on the magnetic structure, NPD data were collected for the SG and SSM series. A representative combined Rietveld refinement of the NPD and PXRD data of the SG sample for $x = 2$ (*i.e.* $SrFe_{10}Al_2O_{19}$) is shown in Figure 7.

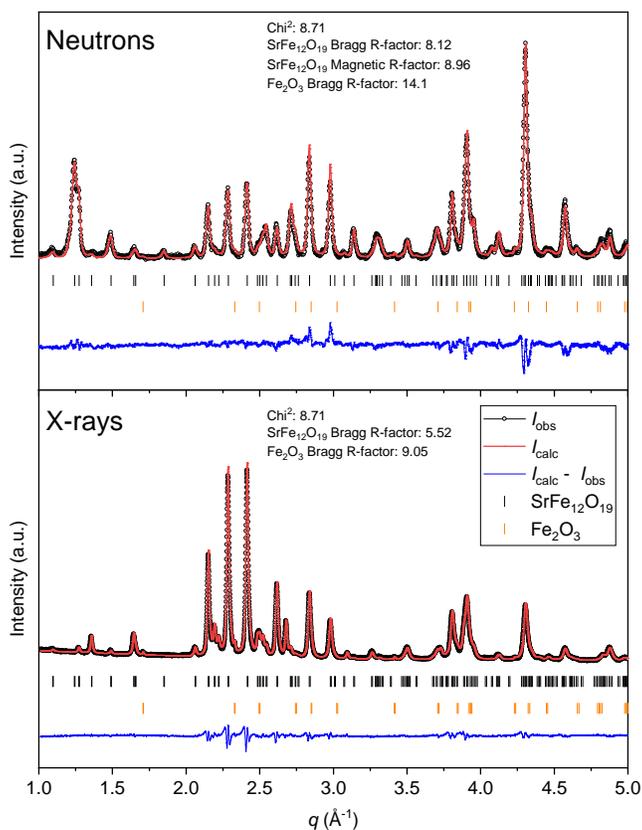

Figure 7: NPD (top) and PXRD (bottom) patterns, and combined Rietveld refinement of SG-synthesized sample with Al substitution of $x = 2$ (*i.e.* $SrFe_{10}Al_2O_{19}$).

To determine which crystallographic sites are occupied by Al in the SG $SrFe_{12-x}Al_xO_{19}$ samples, the following refinement strategy was used: Each of the Fe Wyckoff sites were allowed to be populated by either Fe or Al. Linear restraints were imposed on all the Fe/Al sites to avoid unphysical over or under population of each of the sites, while keeping the overall nominal stoichiometric ratio between Fe and Al within the structure fixed (i.e. 12:0, 11:1, 10:2 and 9:3 for $x = 0$, 1, 2, and 3, respectively). The total amount of aluminum was initially distributed evenly on each of the Fe sites. The relative occupation fraction of each site was then refined, allowing the overall amount of Al to be distributed freely among the different sites. The refined site occupation fractions of Al in each of the crystallographic Wyckoff sites are shown in Figure 8. The remaining occupation fraction is occupied by Fe. The numeric values, atomic coordinates, isotropic displacement parameter and refined magnetic moment of SG sample with $x = 2$ are given in Table 2. Equivalent tables of the remaining samples can be found in the Supporting Information. The refinements clearly indicate that Al preferentially occupies the octahedral 2*a* site, $(2a)_{Oh}$. This site preference is followed by the $(12k)_{Oh}$ site. These two sites are significantly more occupied by Al than the remaining three, and both have the atomic magnetic dipolar moment along the predominant direction (here considered as magnetic moment pointing up).

For the SSM samples, a slightly different refinement strategy was followed. Given the very subtle variation of unit cell parameters and magnetic properties in the SSM series, the assumption of nominal Al insertion in the structure was deemed unfeasible. Thus, to obtain an estimated amount of Al insertion, the relative change in unit cell parameters of the SSM samples with respect to the unsubstituted SSM sample was used. For this, the relative change in both, *a*- and *c*-axes was calculated, and the average value of both was compared to that of the SG series, assuming that the obtained relative change in the SG series lattice parameters corresponds to the expected change for the nominal Al insertion (*i.e.*, $x = 1$, $x = 2$ and $x = 3$), and that an equal relative change in unit cell axes would take place for the SSM and SG series. The amount of Al, *i.e.*, $x$ in $SrFe_{12-x}Al_xO_{19}$, was then estimated to be approx. 0.2, 0.4 and 0.3 for the Al-substituted SSM samples with nominal $x = 1$, 2 and 3, respectively. Once $x$ was estimated, a similar refinement strategy to that of the SG series was followed. The corresponding amount of Al was evenly distributed among the five Fe Wyckoff sites, which were allowed to be populated by either Fe or Al. However, no restraints were imposed to maintain the overall ratio between Fe and Al, allowing $x$ to differ from the aforementioned estimation. The refinements initially gave rise to negative or negligible Al occupation in all Fe sites except for the octahedral 2*a* site (represented in blue in Figure 8b). Notably, this is the same site found to be preferred by Al in the SG-synthesized samples (see Figure 8). Therefore, for the final refinements of the SSM series, the Al content was fixed to zero on all sites except the octahedral 2*a* site. The refinements returned a value of $x$ in the three substituted SSM samples of 0.4, 0.6 and 0.45, which translates into the $(2a)_{Oh}$ site having a 5%, 7% and 5.5% occupation of Al respectively (see Tables S2-S4 in Supporting Information). It should be noted that this amount of Al is very small, and similar quality refinements could be obtained considering no Al present in the samples.

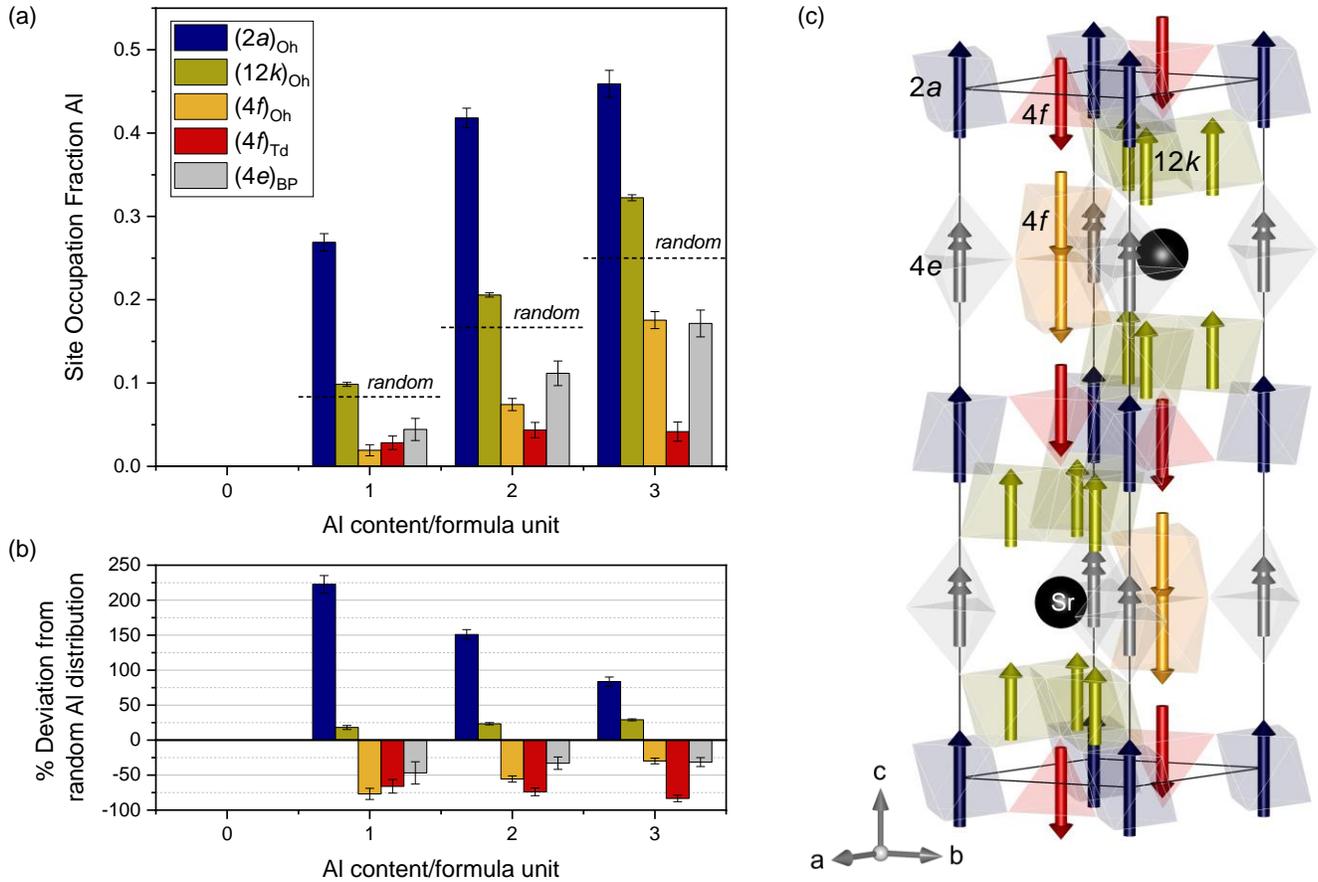

Figure 8: a) Refined site occupation fraction of Al cations in each of the 5 crystallographic sites for the SG-synthesized samples. b) Percentage deviation from random distribution of Al cations in the 5 sites. c) Illustration of the crystal and magnetic structure of $SrFe_{12}O_{19}$ with the crystallographic sites color-coded in accordance with a) and b).

Table 2: Refined crystallographic parameters for aluminum substituted SrFe$_{12-x}$Al$_x$O$_{19}$ ($x = 2$) synthesized by the SG method. The Sr and O atoms were refined with full occupancy. The positive/negative sign in the refined magnetic moment indicates the direction of the magnetic moment along the z-axis in that site. Where no errors are given, the parameters were fixed in the last refinement cycle. Equivalent tables for the remaining samples are provided in Supporting Information.

| SG – SrFe$_{12-x}$Al$_x$O$_{19}$ ($x = 2$) | | | | | | | |
|---|---|---|---|---|---|---|---|
| Space group | *P*6$_3$/*mmc* (No. 194) | | | | | | |
| a=b | 5.841896(8) Å | | | | | | |
| c | 22.9321(4) Å | | | | | | |
| α=β | 90 ° | | | | | | |
| γ | 120 ° | | | | | | |
| | Site occupation fraction (%) | | | | | B$_{iso}$ | R$_z$ |
| Wyckoff site | Fe | Al | x | y | z | (Å$^2$) | μ$_B$ |
| Sr (2*d*) | - | - | 1/3 | 2/3 | 3/4 | 1.84(8) | - |
| Fe1 (2*a*)$_{Oh}$ | 58(1) | 42(1) | 0 | 0 | 0 | 1.80(4) | 5.0(5) |
| Fe2 (12*k*)$_{Oh}$ | 74.4(3) | 20.6(3) | 0.1680(4) | 0.3359(8) | -0.10841(7) | 1.80(4) | 3.0(1) |
| Fe3 (4*f*)$_{Oh}$ | 92.6(7) | 7.4(7) | 1/3 | 2/3 | 0.1892(1) | 1.80(4) | -3.8(2) |
| Fe4 (4*f*)$_{Td}$ | 95.6(9) | 4.4(9) | 1/3 | 2/3 | 0.0267(1) | 1.80(4) | -3.9(2) |
| Fe5 (4*e*)$_{BP}$ | 89(1) | 11(1) | 0 | 0 | 0.249(7) | 1.80(4) | 2.3(3) |
| O1 (4*e*) | - | - | 1/3 | 2/3 | 0.1516 | 0.79(5) | - |
| O2 (4*f*) | - | - | 0 | 0 | -0.0552 | 0.79(5) | - |
| O3 (6*h*) | - | - | 0.1828 | 0.3656 | 1/4 | 0.79(5) | - |
| O4 (12*k*) | - | - | 0.1563 | 0.3126 | 0.0524 | 0.79(5) | - |
| O5 (12*k*) | - | - | 0.5051 | 0.0102 | 0.151 | 0.79(5) | - |

This is the first experimental structural evidence of the preferred Al positions when substituted into the hexaferrite. The results are in good agreement with the theoretical calculations by V. Dixit *et al.*,[24] which predict the (2*a*)$_{Oh}$ and (12*k*)$_{Oh}$ to be the preferred substitution sites for Al. However, according to their calculations, which are restricted to an Al content of $x = 0.5$ and $x = 1$, the preference for Al occupying the (2*a*)$_{Oh}$ or (12*k*)$_{Oh}$ site varies with annealing temperature during synthesis. For the synthesis temperature employed here in the SG and SSM samples (925 °C and 790 °C), their calculations predict occupation of the (12*k*)$_{Oh}$ site to be dominant compared to (2*a*)$_{Oh}$, and to increase from $x = 0.5$ to $x = 1$. Our results do show that the relative occupation of Al in the (12*k*)$_{Oh}$ site increases with increasing Al content. However, the (2*a*)$_{Oh}$ site remains the dominant substitution site for all compositions, in contrast to the calculations by V. Dixit *et al.* for such synthesis temperatures. According to their calculations, the probability of the $Al^{3+}$ ion occupying the $(4f)_{Oh}$, $(4f)_{Td}$ or $(4e)_{BP}$ sites is much lower compared to that of the other two sites. This is also what we observe here, both for the SG and the SSM series. Nonetheless, the calculations by V. Dixit *et al.* indicate the occupation of the $(4f)_{Oh}$, $(4f)_{Td}$ and $(4e)_{BP}$ sites to be negligible, while the refinements of the SG series show that some Al substitution takes place in all sites for the three compositions. The $(4e)_{BP}$ is the next most populated site after $(2a)_{Oh}$ and $(12k)_{Oh}$ for values of $x = 1$. However, as the concentration of Al increases, so does the Al occupation of the $(4f)_{Oh}$ site, reaching equal occupation to that of $(4e)_{BP}$ for $x = 3$. The relative affinity of Al for the different sites and its variation with increasing doping can be clearly evaluated when calculating the deviation from random occupation of Al on each site (Figure 8b). It is evident from Figure 8b that Al has a particular affinity for the $(2a)_{Oh}$ and $(12k)_{Oh}$ sites, and therefore the occupation of Al in those sites is higher compared to a random distribution. The three remaining sites, *i.e.* $(4f)_{Oh}$, $(4f)_{Td}$, and $(4e)_{BP}$, are "avoided" by Al, and are thus being underpopulated with respect to a random occupation. However, as the Al content increases, the occupation of Al on the $(2a)_{Oh}$ and $(12k)_{Oh}$ sites is reduced, and the occupation of $(4f)_{Oh}$ increases. These results clearly display how the relative affinity for the $(4f)_{Td}$, and $(4e)_{BP}$ sites (represented in red and grey respectively in Figure 8) stays unaffected (within the error) by the increase of total Al content in the structure, and the interplay of Al site occupation takes place between the remaining three sites, which are all octahedrally coordinated sites, with moment pointing up in $(2a)_{Oh}$ and $(12k)_{Oh}$ or down in $(4f)_{Oh}$.

From the Rietveld refinements of the NPD patterns, the magnetic moments of the five different $Fe^{3+}$ crystallographic sites were extracted. The refined absolute values of the moments for the SSM and SG samples with increasing $Al^{3+}$ content are shown in Figure 9a and b, respectively, and numerical values are given in Table 2 and Tables S1-S8 in Supporting Information.

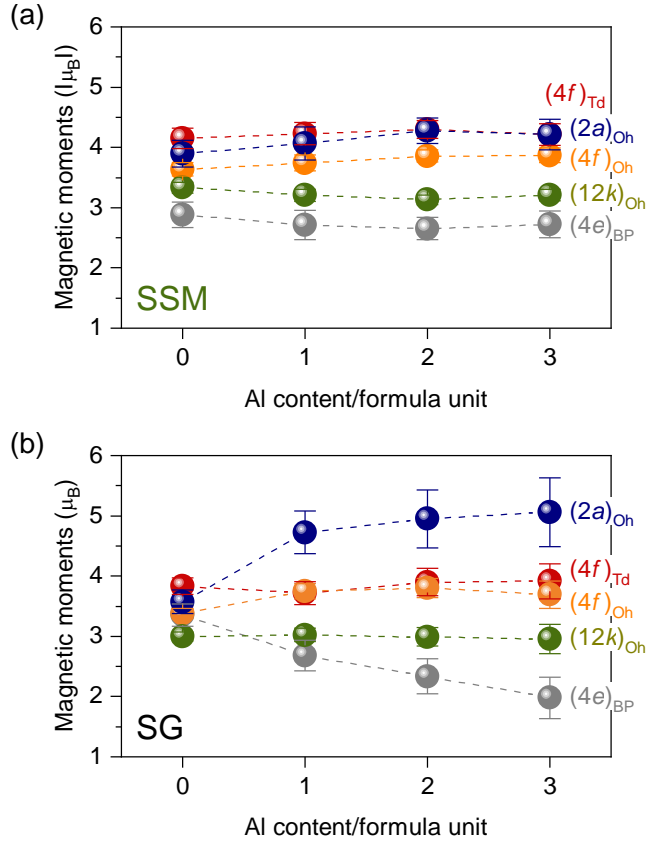

Figure 9: Refined magnetic moments of $Fe^{3+}$ on the five different crystallographic $Fe^{3+}$ sites with increasing aluminum content, for samples synthesized by (a) SSM and (b) SG methods.

The refined magnetic moments of the SSM samples remain practically unchanged with increasing Al content. These results confirm, once again, a very minor degree of Al substitution into the $SrFe_{12}O_{19}$ crystallites by the solid-salt-matrix synthesis method. In the SG samples, the magnetic moment of $Fe^{3+}$ in three crystallographic sites $(12k)_{Oh}$ (green), $(4f)_{Oh}$ (orange) and $(4f)_{Td}$ (red) remains largely unaffected by partial substitution of Al. However, the average magnetic moment of $Fe^{3+}$ in the $(2a)_{Oh}$ site (blue) increases, while it decreases in the $(4e)_{BP}$ site (gray), showing a clear impact of the Al substitution on the atomic magnetic structure. As shown earlier (see Figure 8), Al has the highest affinity for the $(2a)_{Oh}$ site (blue) and thus to maintain the long-range magnetic structure, a stronger moment may be required by the remaining Fe on the $(2a)_{Oh}$ site. Although much more subtle, a slight increase in the magnetic moment of the $(2a)_{Oh}$ site is also observed in the SSM series, where refinements indicated it to be the only site with Al substitution. In contrast, the neighboring bipyramidal $(4e)_{BP}$ site (gray), where only moderate Al substitution is observed (see Figure 8), exhibits a decrease in magnetic moment that may be attributed to the lack of stabilizing moments (*i.e.* higher Al substitution) on the corner-sharing $(12k)_{Oh}$ site, through which super-exchange interactions take place to give rise to the antiferromagnetic long-range structure. This site, $(4e)_{BP}$, which is located in the atomic layer where a Sr atom replaces an oxygen atom, plays a key role in the hexaferrite structure. The spin contribution from the iron atom in this site is the major contributor to the uniaxial anisotropy of the compound.[1, 35] However, due to the proximity of the Sr atom, the $Fe^{3+}$ atom in the bipyramidal site has fewer Fe-O-Fe superexchange connections with neighboring $Fe^{3+}$ atoms, and its magnetic moment is hence highly affected by modifications in the composition of the surrounding Fe sites. Our previous studies on $SrFe_{12}O_{19}$ nanoplatelets also showed how the magnetic moment in the $(4e)_{BP}$ site was highly affected by extreme reduction in the platelet thickness.[6] This also explains that the SSM samples consistently has the lowest magnetic moment on the $(4e)_{BP}$ site as these crystallites are only ~23 nm thick.

The intrinsic crystallographic magnetization for each sample can be extracted from the Rietveld refinements of the NPD data taking into account the refined magnetic moment of $Fe^{3+}$ per site, the multiplicity of each site and the site occupation fraction of Fe on each site. A comparison between the measured ($M_s$, $M_r$) and refined ($M_{NPD}$) magnetization values of the SSM and SG series is shown in Figure 10.

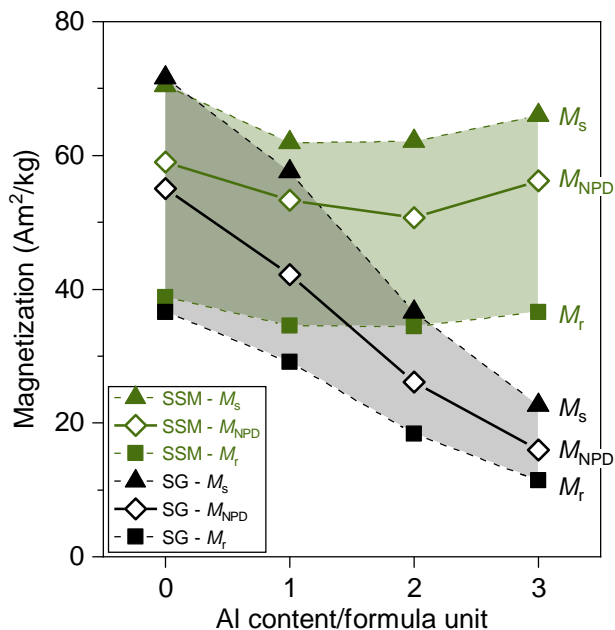

Figure 10: Measured saturation magnetization ($M_s$), remanent magnetization ($M_r$) and calculated magnetization from the refined magnetic moments of the NPD data ($M_{NPD}$) for both, the SG and SSM series, as function of increasing Al doping. Error bars within the size of the symbols.

For the SSM samples (green in Figure 10), a slight decrease is observed in both the measured and refined magnetization values for aluminum substitution. However, very similar values are obtained for $x = 1-3$. This is consistent with the Rietveld results, indicating that a very small amount of Al is indeed substituted in the structure of the SSM samples, but it is very minor, it does not correspond to the nominal Al content, and it does not vary much regardless of an increasing amount of Al being added during the synthesis process. As confirmed by the agreement in the trend between the measured and refined magnetization values (Figure 10), the small decrease in the attained magnetization can be explained by the minor amount of non-magnetic $Al^{3+}$ cations effectively substituted in the $(2a)_{Oh}$ site.

For the SG series (black in Figure 10), the calculated magnetization extracted from the Rietveld refinement of the NPD data shows a continuous decrease in saturation magnetization with increasing Al content, which is consistent with the observed trend in the measured magnetic data. The reduction in the total magnetization of the sample with increasing Al content is due to the non-magnetic nature of $Al^{3+}$ cations and their specific site occupation within the crystal structure.

As shown in Figure 10, the $M_{NPD}$ value is found to be precisely between that of the saturation and remanent magnetization values obtained from VSM measurements, as expected. The measured saturation $M_s$ should be equivalent to the refined magnetization, if NPD data were collected at 0 K. Here, however, the refined magnetization is obtained from room temperature NPD data. At room temperature, and without any applied magnetic field, the atomic magnetic dipolar moment will precess due to thermal fluctuations. This will reduce the projection of the magnetic moment vectors onto the $c$-axis with respect to the fully aligned moments and therefore reduce the "observed" scattering from the atomic magnetic dipolar moments.

The excellent agreement between refined and measured magnetization values validates the results obtained in the NPD refinements, which indicate a preferred site occupation of the $Al^{3+}$ cations in the $(2a)_{Oh}$ and $(12k)_{Oh}$ Wyckoff sites, followed by $(4e)_{PB}$ and $(4f)_{Oh}$. Given that the three most substituted sites correspond to sites with the magnetic moment pointing up (main magnetization direction), substitution of $Fe^{3+}$ by the non-magnetic $Al^{3+}$ onto those sites explains the observed reduction in the intrinsic magnetization of the compounds. The increase in coercivity can be explained by (and be a consequence of) the reduction in the saturation magnetisation, if it is assumed that the magnetocrystalline anisotropy ($K_1$) is the same for all samples, given that the coercivity is related to the saturation magnetisation by the relation $H_c = 2K_1/M_s$.

## Conclusions

Partial substitution of Al into the $SrFe_{12}O_{19}$ structure was attempted by three synthesis methods, namely autoclave (AC), citrate sol-gel (SG) and solid-salt-matrix (SSM) synthesis. While no (or negligible) effective Al substitution was observed in the AC or SSM samples, successful Al substitution was attained in the SG samples, leading to a decrease of the unit cell axes and an increase of crystallite size with increasing Al content. Combined Rietveld refinements of NPD and PXRD data revealed that the non-magnetic aluminum atoms predominantly occupy the sites with the magnetic moment pointing up; i.e. $(2a)_{Oh}$, $(12k)_{Oh}$ and $(4e)_{BP}$, but as the Al content increases, the $(4f)_{Oh}$ site whose magnetic moment is pointing down, also becomes partially populated by Al. As a consequence of Al predominantly occupying sites with the magnetic moment pointing up, the net magnetic moment of the compound is diminished. The macroscopic magnetization of the samples is reduced with increasing Al doping, which is in excellent agreement with the refined magnetic moments and Al site occupation fractions, as well as with previously reported theoretical calculations. Notably, very high coercivity values are obtained by Al substitution, reaching a maximum of 830(21) kA/m (10.4 kOe) for $SrFe_9Al_3O_{19}$. Not only does this value exceed that of the un-substituted $SrFe_{12}O_{19}$ by ~71% but is comparable to the coercivity of $Nd_2Fe_{17}B$ magnets.

## Acknowledgements

This work was financially supported by the Independent Research Fund Denmark – METEOR (1032-00251B), the Danish Center for Synchrotron and Neutron Science (DanScatt), and the Spanish Ministry of Science and Innovation through Grant TED2021-130957B-C52 (NANOBOND) funded by MCIN/AEI/10.13039/501100011033 and by the "European Union NextGenerationEU/PRTR". This project has received funding from the European Union's Horizon Europe research and innovation programme under project No 101063369 (OXYPOW). Matilde Saura-Múzquiz gratefully acknowledges the financial support from the Comunidad de Madrid, Spain, through an "Atracción de Talento Investigador" fellowship (2020-T2/IND-20581). The authors gratefully acknowledge the beamtime granted by the Australian Centre for Neutron Scattering, Australian Nuclear Science and Technology Organisation. Affiliation with the Center for Integrated Materials Research (iMAT) at Aarhus University is gratefully acknowledged. Jakob Voldum Ahlburg and Frederik Holm Gjørup are thanked for their assistance with the VSM data extraction and fruitful discussions.